\documentclass[floatfix,nofootinbib,amsmath,amssymb,prd,twocolumn,10pt,aps]{revtex4-2}
\usepackage{mathrsfs}
\usepackage[x11names]{xcolor}
\usepackage{graphicx}
\usepackage{dcolumn}
\usepackage{bm}
\usepackage{soul}
\usepackage{hyperref}
\usepackage{orcidlink}
\usepackage[skip=5pt, indent=10pt]{parskip}
\usepackage{tabularray}
\usepackage[normalem]{ulem}
\usepackage{physics} 
\usepackage{subcaption}

\hypersetup{
  pdftitle={Sidereal Signals of Lorentz Violation in Atmospheric Neutrinos},
  pdfsubject={SME, neutrinos, Lorentz and CPT violation, Matter Efects},
  pdfauthor={Ioannou-Nikolaides, Hilding-Nørkjær, Koskinen, Stuttard},
  pdfstartview={FitH},
  colorlinks=true,
  linkcolor=Blue1!60!black,
  citecolor=Green1!50!black,
  urlcolor=Blue1!70!black
}

\usepackage{tikzpagenodes}

\begin{document}

\title{Interplay of Lorentz Invariance Violation and Earth's Matter Potential in High-Energy Neutrinos}

\author{Simon Hilding-Nørkjær\,\orcidlink{0009-0001-9091-3254}\thanks{Authors are listed in alphabetical order and contributed equally to this work.}}
\email{simon.hilding@nbi.ku.dk}
\affiliation{Niels Bohr Institute, University of Copenhagen, Blegdamsvej 17, 2100 Copenhagen, Denmark}

\author{Johann Ioannou-Nikolaides\,\orcidlink{0000-0003-2486-1588}\thanks{Authors are listed in alphabetical order and contributed equally to this work.}}
\email{johann.nikolaides@nbi.ku.dk}
\affiliation{Niels Bohr Institute, University of Copenhagen, Blegdamsvej 17, 2100 Copenhagen, Denmark}

\author{D.~Jason Koskinen\,\orcidlink{0000-0002-0514-5917}\thanks{Authors are listed in alphabetical order and contributed equally to this work.}}
\affiliation{Niels Bohr Institute, University of Copenhagen, Blegdamsvej 17, 2100 Copenhagen, Denmark}

\author{Thomas Stuttard\,\orcidlink{0000-0001-7944-279X}\thanks{Authors are listed in alphabetical order and contributed equally to this work.}}
\affiliation{Niels Bohr Institute, University of Copenhagen, Blegdamsvej 17, 2100 Copenhagen, Denmark}

\begin{abstract}
Searches for Lorentz invariance violation (LIV) in the neutrino sector have traditionally focused on non-standard neutrino oscillations induced by LIV in vacuum. In this work, we study anisotropic LIV in matter. First, we review vacuum LIV phenomenology, explaining the energy and direction dependence of sidereal modulations for anisotropic coefficients in the Standard Model Extension. We then demonstrate that for high-energy neutrinos, the interplay between anisotropic LIV operators and the Earth's matter potential produces distinct, observable signatures absent in the vacuum case. We identify a crossover regime where the energy-dependent LIV Hamiltonian becomes comparable to the matter potential, leading to strong interference effects. By analyzing the propagation of neutrinos through a realistic Earth model, we establish three key phenomenological consequences: (1) direction-dependent resonant enhancements of oscillation probabilities, (2) a macroscopic breakdown of neutrino-antineutrino symmetry for CPT-even operators, and (3) a significant increase of the $\nu_\tau$ flux due to LIV-driven injection of high-energy neutrinos into the $\tau$ regeneration cycle. 
These results highlight that accounting for the interplay between LIV and matter is essential for future LIV searches at large-scale neutrino telescopes.
\end{abstract}

\maketitle
\section{Introduction}

Neutrino oscillations describe the experimentally verified process of neutrinos changing into each other while they propagate. In the Standard Model there are three different neutrino flavors $\nu_e$, $\nu_\mu$ and $\nu_\tau$ corresponding to the three lepton flavors produced alongside when neutrinos are produced. 
Yet the propagation of neutrinos occurs in mass eigenstates, which are the eigenvectors of the neutrino Hamiltonian. The mixing between the flavor and mass eigenstates arises from the nonzero mass differences between the mass eigenstates and is described by the PMNS mixing matrix\,
\cite{Pontecorvo:1967fh,makiRemarksUnifiedModel1962,bilenkyLeptonMixingNeutrino1978}. The mass differences between the three mass eigenstates are called the \textit{solar} mass difference $\Delta m^2_{12}$, and the \textit{atmospheric} mass difference $\Delta m^2_{3\ell}$. These names are derived from the sources of the neutrinos in the experiments that first measured their oscillations\,\cite{Giunti2007,Super-Kamiokande:1998kpq,collaborationDirectEvidenceNeutrino2002}.

The phenomenon of neutrino oscillations is firmly established, and a large number of experiments have provided precise measurements and constraints on the fundamental oscillation parameters\,\cite{ParticleDataGroup:2022pth,Esteban:2020cvm}.
Since oscillations arise from accumulated phase differences during propagation, they are highly sensitive to small perturbations of the neutrino Hamiltonian. Due to this sensitivity neutrinos are ideal probes of new physics\,\cite{arguellesWhitePaperNew2020}, including matter-induced potentials\,\cite{smirnovMSWEffectMatter2005}, nonstandard interactions\,\cite{farzan_tortolaNeutrinoOscillationsNonStandard2018,collaborationStrongConstraintsNeutrino2022}, decoherence\,\cite{lisiProbingPossibleDecoherence2000,Stuttard:2020qfv}, and effects associated with spacetime symmetries\,\cite{Kostelecky:2003LIVCPTVNeutrinos}.

The fundamental group of spacetime symmetries consists of Lorentz transformations and translations. The invariance of the laws of physics under spacetime symmetries and the combination of charge conjugation (C), parity inversion (P), and time reversal (T) form the fundamental principles of quantum field theory\,\cite{weinbergQuantumTheoryFields2013}.
Neutrinos are particularly well-suited to test the violation of Lorentz-invariance (LIV) and invariance under CPT symmetry (CPTV), as neutrinos only interact through the weak force and gravity. The long distances over which neutrinos can propagate undisturbed permit even the smallest perturbations to accumulate, resulting in an observable effect.

An effective field theory that includes all possible forms of LIV and CPTV is the Standard Model Extension (SME)\,\cite{Colladay:1996iz,Colladay:1998fq,Kostelecky:2003fs}\footnote{While CPT violation implies Lorentz violation in local quantum field theories~\cite{GreenbergCPTviolationImpliesLIV}, non-local models that do not imply Lorentz violation have been proposed~\cite{Chaichian:2011fc}. The operators introduced in the SME are all Lorentz invariance violating and can be CPT violating. We will therefore refer to the effects of the SME operators on neutrinos as Lorentz-invariance violating (LIV) to encompass scenarios where only Lorentz invariance is violated or both symmetries are broken.}. This model-independent formalism has been used extensively to constrain the rich phenomenology of LIV in the neutrino sector\,\cite{kosteleckyLorentzCPTViolation2004,kosteleckyDataTablesLorentz2011}.

SME operators modify the effective neutrino Hamiltonian, and manifest as modified neutrino oscillations that can exhibit a nonstandard energy dependence, rather than the $L/E$ dependence expected from standard oscillations\,\cite{Kostelecky:2003LIVCPTVNeutrinos,kosteleckyNeutrinosLorentzviolatingOperators2012}. They can also be direction-dependent.
The direction-dependence can be intuitively understood as an aether-like background tensor field. As the detectors rotate relative to the background field's preferred direction, a time-dependence is introduced to the nonstandard oscillations induced by the SME operators\,\cite{Kostelecky:2003LIVCPTVNeutrinos,abbasiSearchQuantumGravity2022,Diaz:2011ia,kosteleckyNeutrinosLorentzviolatingOperators2012}. 

In the standard Sun-centered inertial frame used for SME analyses\,\cite{kosteleckyDataTablesLorentz2011,bluhmClockComparisonTestsLorentz2002,bluhmProbingLorentzCPT2003}, the background field is constant. However, the Earth's rotation changes the intersection of the neutrino's propagation path with regions of the Earth's interior, causing the matter density profile along a path to vary over time. Consequently, a comprehensive LIV study requires a careful consideration of the interplay between the neutrino masses, the SME operators and the matter potential experienced by neutrinos propagating through the Earth.

To conduct our study on anisotropic LIV in matter, we developed a software implementation of anisotropic Lorentz- and CPT-violation effects in neutrino oscillations. This implementation is publicly available as a dedicated branch of the {\ttfamily C++} based {\ttfamily nuSQuIDS} package\,\cite{Arguelles:2021twb} on \textit{GitHub}\,\cite{stuttard_nusquids_bsm}. Our implementation specifically handles the propagation of neutrinos through varying matter potentials for various SME operators. It is meant to be extendable and we provide {\ttfamily DEIMOS}\,\cite{stuttard_deimos}, a simple Python-based package that acts as an easy-to-use wrapper around {\ttfamily nuSQuIDS}.

\section{Lorentz Invariance Violating Neutrinos in the Standard Model Extension}\label{ch1}


The evolution of neutrino states during propagation in vacuum is governed by the standard neutrino Hamiltonian
\begin{align}
    H_0= \frac{1}{2E}U m^2 U^\dagger,
    \label{eq: standard neutrino Hamiltonian}
\end{align}
where $m^2$ is the diagonal mass matrix, $U$ is the PMNS matrix and $E$ is the neutrino's energy. To model the evolution of neutrinos in matter, the matter potential $V$ is added to Eq.\,\eqref{eq: standard neutrino Hamiltonian}. 

The full Hamiltonian accounting for all possible LIV violation terms involving fermion bilinears then takes the form 
\begin{align}
    H= H_0 + V + H_\text{SME},
    \label{eq: full Hamiltonian}
\end{align}
where the SME introduces direction-, energy-, and flavor-dependent terms. For neutrino–neutrino and antineutrino–antineutrino sectors, the renormalizable contributions are
\begin{align}
    H_\text{SME} = \frac{1}{E}\Bigl(\widehat{a}_{L}^{(3)\alpha} p_\alpha-\widehat{c}_{L}^{(4)\alpha\beta} p_\alpha p_\beta\Bigr),
    \label{eq: HLIV}
\end{align}
where $p_\alpha$ is the momentum 4-vector, which for relativistic neutrinos propagating in the direction $\widehat{p}$ can be approximated by $p_\alpha\approx E\mqty(1,&-\mathbf{\widehat{p}})$\,\cite{kosteleckyLorentzCPTViolation2004}. The exponents $(3)$ and $(4)$ indicate the dimensionality of the operators, while the subscript $L$ indicates that the operator applies to left-handed neutrinos specifically.
In this work we present results for the renormalizable CPT-odd operator $\widehat{a}_{L}^{(3)}$ and the CPT-even operator $\widehat{c}_{L}^{(4)}$ explicitly and discuss how these results generalize to higher-dimensional operators. All operators are $3\times3$ matrices in the flavor basis or mass basis and act as a background tensor field that modifies neutrino propagation depending on their energy, direction, and particle-antiparticle nature. 

Since the SME operators act as a fixed background field in an inertial frame, their directional dependence can be expressed using spherical harmonics\,\cite{kosteleckyNeutrinosLorentzviolatingOperators2012}.
\begin{align}
    (\widehat{c}_L^{(4)\alpha\beta})_{ij}\,p_\alpha\,p_\beta =  \sum_{l=0}^2\sum_{m=-l}^l E^{4-2}\,Y_{lm}(\widehat{p})(\widehat{c}_L^{(4)})^{lm}_{ij}
    \label{eq: Spherical decomposition aL3}
\end{align}
and analogously for $\widehat{a}_{L}^{(3)\alpha}$. The indices $i$, $j$ label the neutrino mass eigenstates, while $Y_{lm}(\widehat{p})$ are the spherical harmonic functions. The maximal degree $l$ grows with the dimensionality of the operator. In the case of $(\widehat{a}_L^{(3)\alpha})_{ij}$ and $(\widehat{c}_L^{(4)\alpha\beta})_{ij}$ it  obeys $0\leq l<2$ and $0\leq l<3$. These parameters can produce fields with monopole and dipole moments in the case of $(\widehat{a}_L^{(3)\alpha})_{ij}$ and additionally quadrupole moments in the case of $(\widehat{c}_L^{(4)\alpha\beta})_{ij}$. \\
Higher dimensional operators are contracted with more momentum 4-vectors. These contractions yield higher degrees of anisotropies and have a stronger energy-dependence, which we discuss in Subsection \ref{sec1}.

To illustrate the key physical consequences of the anisotropic SME operators with minimal complexity, we focus on the dipolar components $\beta=x,y,z$ of $(\widehat{c}_L^{(4)t\beta})_{33}$ in the mass basis. In selected examples we also include the coefficient $(\widehat{a}_L^{(3)\alpha})_{33}$ to highlight the qualitatively different signatures produced by operators that are CPT-odd, and have a different energy scaling. 

The coefficients $(\widehat{c}_L^{(4)t\beta})_{33}$, and $(\widehat{a}_L^{(3)\alpha})_{33}$ perturb the propagation of the third mass eigenstate and are strongest for neutrinos traveling along the $X$, $Y$, and $Z$ coordinate axes of the Sun-centered frame, respectively.
Each breaks the rotational symmetry of standard neutrino oscillations described by Eq.\,\eqref{eq: standard neutrino Hamiltonian} and leads to anisotropic effects, which we discuss in Subsection \ref{sec2}.
The choice of the specific SME coefficients is illustrative rather than limiting. It produces clear modifications to $\nu_\mu\leftrightarrow\nu_\tau$ oscillations, directly tying LIV effects to the atmospheric sector, where the matter potential becomes relevant due to the Earth-crossing baselines.

In previous experimental studies isotropic models have been popular and are well constrained~\cite{collaborationNeutrinoInterferometryHighPrecision2018, abbasiSearchQuantumGravity2022, collaborationTestLorentzInvariance2015,KM3NeT:2026kuj}. The relationship between existing constraints on isotropic coefficients and anisotropic SME operators depends on the exact search strategy deployed in the isotropic analyses. We discuss this specifically for the isotropic analysis using atmospheric neutrinos \cite{collaborationNeutrinoInterferometryHighPrecision2018}, closest to our study. 
The isotropic ($l=0$) components of the $\widehat{c}_L^{(4)\alpha\beta}$ operator have been tightly constrained at the level of $\sim10^{-28}$, by contrasting upward-going and near-horizontal fluxes to exploit the difference in propagation length in~\cite{collaborationNeutrinoInterferometryHighPrecision2018}. These constraints cannot, however, be straightforwardly interpreted as bounds on the anisotropic ($l\geq 1$) components explored here. The flux-ratio measurement used in ~\cite{collaborationNeutrinoInterferometryHighPrecision2018} is sensitive to the projection of the LIV background field onto the particular geometric axis defined by the analysis. A field aligned with that axis is efficiently constrained, while the sensitivity degrades for any other direction. The $\sim10^{-28}$ bound therefore applies to LIV operators whose signal is maximally expressed along the chosen comparison axis, not to all orientations of an anisotropic field. Dedicated sidereal bounds on the dipolar $(\widehat{c}_L^{(4)t\beta})_{33}$ coefficients studied here come from Refs.\,\cite{collaborationSearchLorentzInvariance2010,icecubeSearchLorentzviolatingSidereal2010}. Following the approach of the Data Tables for Lorentz and CPT Violation\,\cite{kosteleckyDataTablesLorentz2011}, where bounds on individual SME coefficients are interpreted with all others set to zero, the parameter values in this work are chosen to render the novel phenomenological signatures clearly visible and are consistent with the most current bounds on the relevant anisotropic components.

This framework allows us to demonstrate how nonstandard energy dependence, directional anisotropies, sidereal modulations, and matter-enhanced effects affect neutrino propagation, all of which we analyze in the following Sections \ref{ch3} and \ref{sec4}.

\section{Vacuum Phenomenology}\label{ch3}

In this section, we examine the impact of LIV operators on the probabilities of atmospheric neutrino oscillations. First, we introduce the resulting non-standard energy dependence of neutrino oscillations, which appear in both the isotropic and anisotropic components of the SME. Then, we discuss direction-dependent effects specific to the anisotropic SME. This analysis establishes a foundation for understanding the consequences of LIV on neutrino oscillations, thereby facilitating the discussion of the complex interplay between LIV and Earth's matter potential, which we discuss in Section \ref{sec4}.

\subsection{Energy Dependence and Scaling}\label{sec1}

\begin{figure}
    \centering
    \includegraphics[width=\linewidth,trim={0.0cm
    0.0cm 0.0cm 1.0cm},clip]{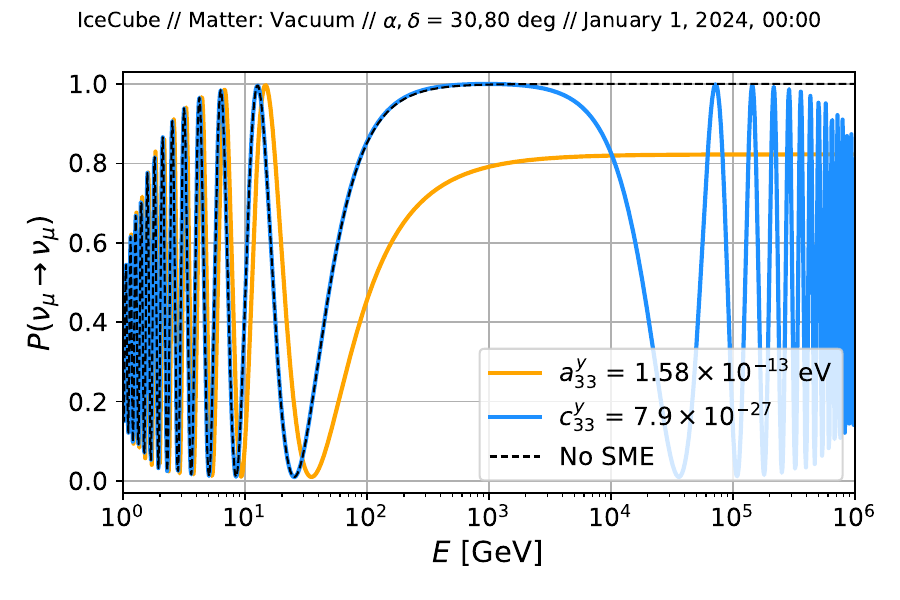}
    \caption{\textbf{Muon neutrino survival probability with and without Lorentz-invariance violation.} The Earth-crossing $\nu_\mu$ survival probability is plotted as a function of energy. Standard oscillations (black) scale as $1/E$, compared to SME-induced oscillations that scale as $E^0$ (orange) and $E^1$ (blue).}
    \label{fig: energy_dep}
\end{figure}
The standard neutrino Hamiltonian in Eq.\,\eqref{eq: standard neutrino Hamiltonian} produces oscillation phases proportional to $\Delta m^2\,L/E$. 
In contrast, the SME terms introduced in Eq.\,\eqref{eq: HLIV} have a different energy-scaling. The contribution from $(a_L^{(3)\alpha})_{ij}$ scales as $E^0$, while the contribution from $(c_L^{(4)\alpha\beta})_{ij}$ scales as $E^1$. Therefore at high energies even weak LIV effects can come to dominate over the mass term contributions. Figure\,\ref{fig: energy_dep} illustrates the scaling behaviors in atmospheric neutrino oscillations for example values of $(a^{(3)y}_L)_{33}$ (orange line) and $(c^{(4)ty}_L)_{33}$ (blue line)\footnote{After converting to the flavor basis, the corresponding value of $(a^{(3)y}_L){\mu\tau}$ is  $0.5\times 10^{-22}$\,GeV and lies within the current bound of $0.6 \times 10^{-22}$\,GeV\,\cite{collaborationSearchLorentzInvariance2010}. Meanwhile, the bound on $(c^{(4)ty}_L)_{\mu\tau}$ is $0.5 \times 10^{-23}$\cite{collaborationSearchLorentzInvariance2010}, which is much looser than the value chosen in Figure\,\ref{fig: energy_dep}. We note that the \textit{isotropic} SME coefficients have been further constrained by several orders of magnitude\,\cite{collaborationNeutrinoInterferometryHighPrecision2018} and they would constrain the picture in Figure\,\ref{fig: energy_dep}. However, the mapping of isotropic limits onto the full anisotropic coefficient space is non-trivial. Therefore, we show these values here to maintain consistency with the anisotropic signatures discussed later}.
Standard oscillations induced by the mass term (black dashed line) become negligible above $\mathcal{O}$(100 GeV) for terrestrial baselines, while the SME-induced oscillations either persist ($(a^{(3)y}_L)_{33}$) or grow with energy ($(c^{(4)ty}_L)_{33}$)\footnote{The blue and orange lines in Figure\,\ref{fig: energy_dep} depict the impact of the mass term and SME Hamiltonian both being non-zero.
For the chosen value of $(a^{(3)y}_L)_{33}$, the mass term dominates at low energies. At high energies, the energy-independent SME-induced contribution dominates, resulting in a fixed $\nu_\mu$ survival probability. 
For $(c^{(4)ty}_L)_{33}$ the picture is similar. At low energies standard oscillations dominate. Compared to the mass-term in Eq.\eqref{eq: full Hamiltonian}, $(c^{(4)ty}_L)_{33}$ has an inverted energy scaling, which manifests in a mirroring of the oscillations.}.

Higher dimensional operators grow even more rapidly with energy and even smaller parameter values become detectable\footnote{This does not necessarily imply an inherent correlation between sensitivity and dimensionality. It is expected that higher order operators are suppressed by the energy scale associated with the violation. If LIV arises from quantum gravity, this energy scale is most naturally expected to be the Planck scale\,\cite{amelino-cameliaQuantumSpacetimePhenomenology2013}.}. 
Large-volume neutrino telescopes operating in the TeV–PeV regime, such as IceCube and KM3NeT, are therefore particularly well suited to constraining these operators.

The growth of higher-dimensional SME operators with energy explains why LIV searches use neutrinos at the highest accessible energies. For neutrino telescopes, astrophysical neutrinos combine the highest observed energies with cosmological propagation distances, making them powerful probes of energy-dependent LIV effects. However, their unknown baselines limit the ability to detect directional or baseline-dependent signatures. Instead, astrophysical neutrinos are used to constrain LIV through altered flavor composition at Earth\,\cite{bustamanteTheoreticallyPalatableFlavor2015,Arguelles:2015dca,abbasiSearchQuantumGravity2022}.\\
Atmospheric neutrinos, by contrast, have a fixed relationship between arrival direction and baseline, enabling studies of directional and baseline-dependent signatures\,\cite{diazPerturbativeLorentzCPT2009,collaborationNeutrinoInterferometryHighPrecision2018}. For this reason, we focus on atmospheric neutrinos for the analysis of directional and matter-dependent LIV effects in the following.

\subsection{Direction Dependence and Sidereal Modulations}\label{sec2}

\begin{figure}
    \centering
    \includegraphics[width=0.95\linewidth,trim={0.0cm
    0.0cm 0.0cm 0.0cm},clip]{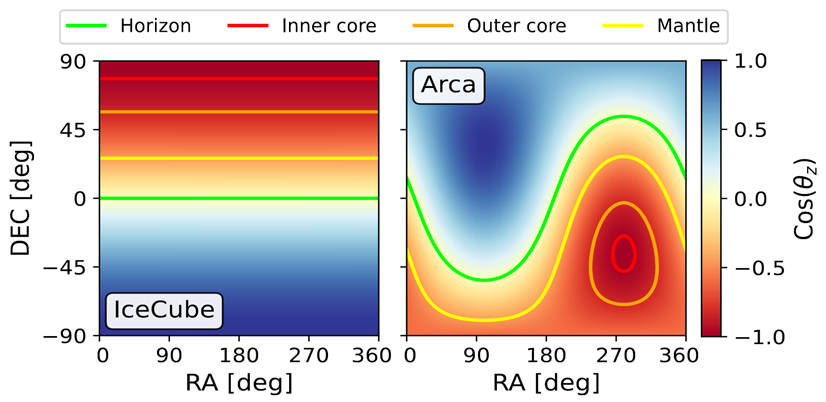}
    \caption{\textbf{Relationship between equatorial and local coordinates.} The cosine of the zenith angle ($\cos\theta_z$), which determines the baseline $L$, is plotted against declination and right ascension. For detectors off the poles, this mapping depends on sidereal time.
    }
    \label{fig: coszen_vs_direction}
\end{figure}

LIV fields in the SME framework are defined in the Sun-centered equatorial frame, where bounds are conventionally reported\,\cite{kosteleckyDataTablesLorentz2011}. To constrain the SME coefficients using atmospheric neutrinos, we must map the local horizontal coordinates (zenith $\theta_z$ and azimuth), which determine the neutrino path length $L$ and matter profile, to the fixed right ascension (RA) and declination (DEC) of the equatorial frame\,\cite{Kostelecky:2003LIVCPTVNeutrinos}. 

\begin{figure}
    \centering
    \includegraphics[width=\linewidth,trim={0.0cm
    0.0cm 0.0cm 0.0cm},clip]{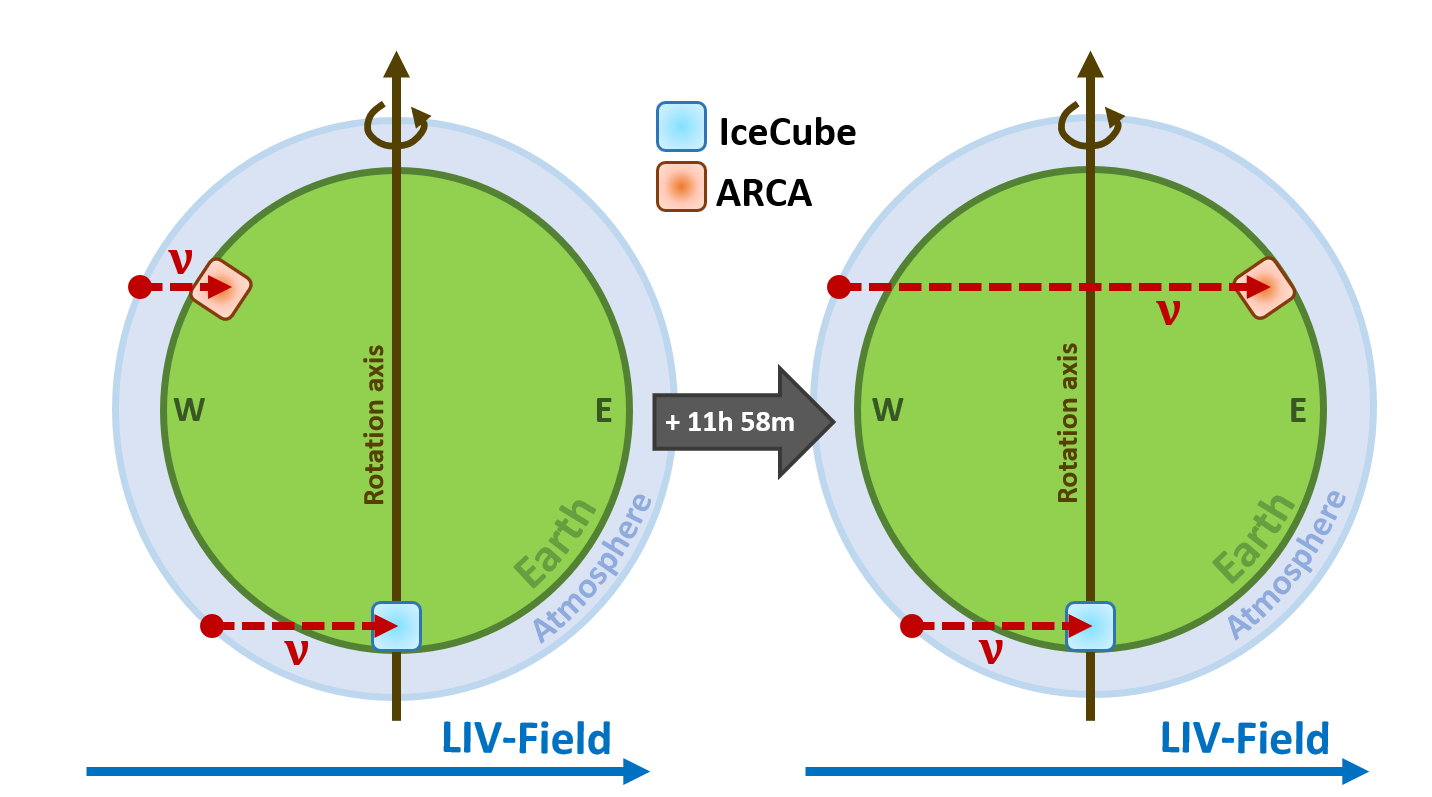}
    \caption{\textbf{Time dependence of the neutrino baseline for a detector located on/off the Earth's rotational axis.} In the celestial equatorial coordinate system, the neutrino baseline is constant for IceCube, while it varies for all other large-scale Cherenkov neutrino telescopes.}
    \label{fig: liv_time_dep_diagram}
\end{figure}

For a detector located at a specific latitude, the Earth's rotation introduces a time-dependence to this mapping. A fixed RA/DEC direction on the sky corresponds to a continuously changing $\theta_z$ over the course of a sidereal day. Consequently, both the path length and the matter profile traversed by neutrinos from that fixed source vary over the course of a sidereal day.
We illustrate the mapping in Figure\,\ref{fig: coszen_vs_direction} for IceCube (South Pole) and KM3NeT/ARCA (Northern Hemisphere) at a fixed time of the sidereal day. \\
For IceCube, on the Earth's rotation axis, the mapping is time-independent and a fixed declination corresponds to a constant zenith angle.\\
In contrast, for a detector off the Earth's rotation axis like KM3NeT/ARCA, the Earth's rotation causes the baseline of a fixed RA/DEC direction to oscillate daily. As illustrated schematically in Figure\,\ref{fig: liv_time_dep_diagram} neutrinos from a specific celestial region may traverse multiple layers of Earth at one time of the sidereal day and much shorter baselines with no matter at another point of time during the day.

This transformation between coordinate frames has an important consequence. Even standard, mass-induced oscillations exhibit a sidereal modulation when analyzed in the equatorial frame because the oscillation phase $\Delta m^2 L/E$ varies over a sidereal day.

\begin{figure}
    \centering
    \includegraphics[width=\linewidth, trim={1cm, 0.5cm, 0cm, 1cm}, clip]{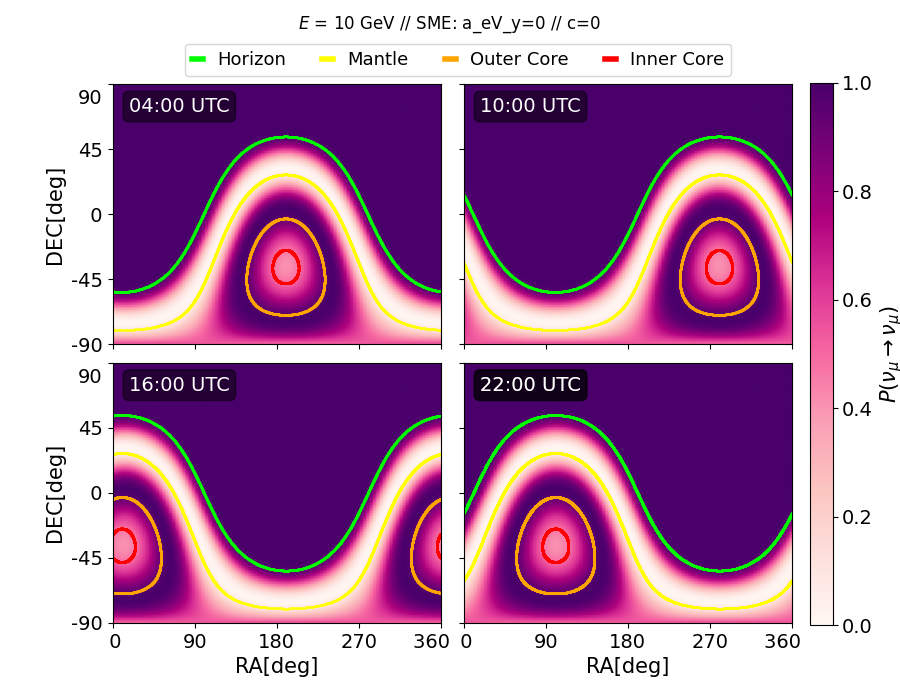}
    \caption{\textbf{Time-dependence of standard neutrino oscillation probabilities in the celestial equatorial reference frame.} We show the sidereal variation of the standard $\nu_\mu$-survival probabilities at the KM3NeT/ARCA detector location at a neutrino energy of $E=10$\,GeV in the equatorial reference frame. We neglect matter effects, but show Earth's main layers for reference.}
    \label{fig: STD variation day signal Toulon}
\end{figure}

To isolate the effects of Lorentz violation, we first establish the baseline behavior of standard mass-induced oscillations in this frame. In Figure\,\ref{fig: STD variation day signal Toulon} we show the sidereal modulation of the $\nu_\mu$ survival probability for KM3NeT/ARCA at $E=10$\,GeV. From the shorter baselines above the horizon with no oscillations, there is a smooth transition to the first oscillation maximum for longer baselines passing through Earth's crust, followed by an oscillation minimum and the increase towards the second oscillation maximum for even longer baselines going through the center of the Earth. 

\begin{figure}[t]
    \centering
    \includegraphics[width=\linewidth, trim={1cm, 0.5cm, 0cm, 1cm}, clip]{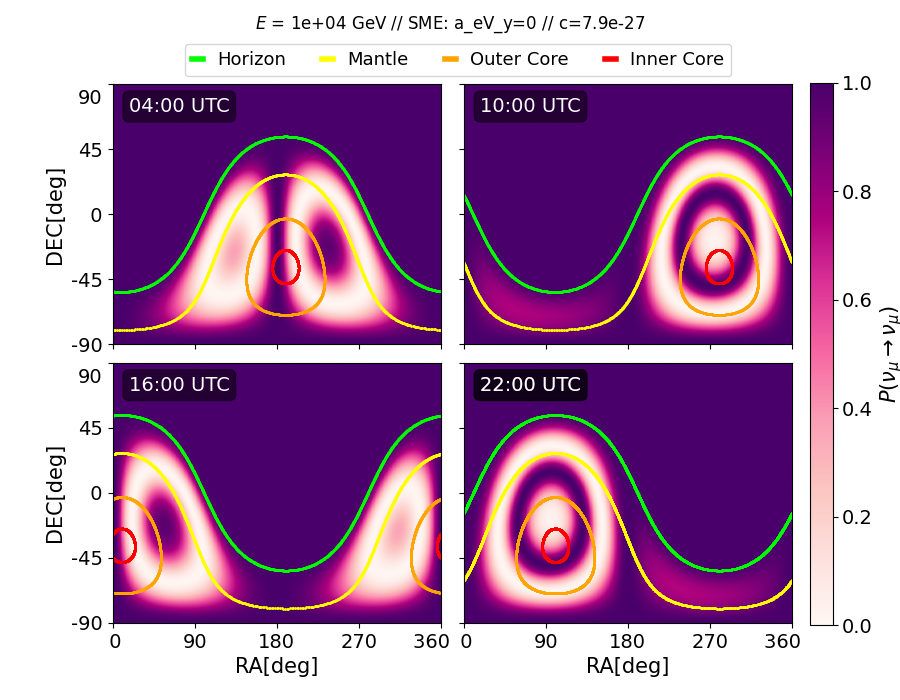}
    \caption{\textbf{Time-dependence of LIV-induced neutrino oscillation probabilities in the celestial equatorial reference frame.} We show the sidereal variation of the vacuum $\nu_\mu$-survival probabilities at the KM3NeT/ARCA detector location for $c^{ty}_{33}=7.9\cdot 10^{-27}$\, at a neutrino  energy of $E=10$\,TeV.} 
    \label{fig: variation day signal Toulon}
\end{figure}

We contrast this with the $\nu_\mu$ survival probability for anisotropic LIV-induced neutrino oscillations at energies ($E=10$\,TeV) well above those at which standard oscillations occur, shown in Figure\,\ref{fig: variation day signal Toulon}. Here, alongside the sidereal modulation there is an additional physical direction dependence. While LIV-induced neutrino oscillations still have the same dependence on the baseline as mass-induced oscillations, comparing Figure\,\ref{fig: STD variation day signal Toulon} and Figure\,\ref{fig: variation day signal Toulon} reveals for LIV-induced oscillations characteristic null directions where the otherwise continuous oscillations are suppressed creating oscillation islands on the RA/DEC plot.

The difference between the mass-induced oscillations and LIV-induced oscillations arises from the anisotropy of the LIV field. Its varying strength for neutrinos originating from different directions introduces an additional direction dependence. The background field is strongest for neutrinos propagating parallel or antiparallel to the field, thereby amplifying their oscillations. Conversely, its effect vanishes completely for neutrinos propagating in directions orthogonal to the field resulting in null bands with no oscillations at all. 

For the specific case of a non-zero dipolar $(\widehat{c}_L^{(4)ty})_{33}$ component in Figure\,\ref{fig: variation day signal Toulon}, the field strength has a sinusoidal dependence on RA with two minima at RA$=0^\circ$ and RA$=180^\circ$, as well as for DEC with two minima at DEC=$-90^\circ$ and DEC=$90^\circ$. 
These null directions persist regardless of the baseline length and are therefore time-independent for a specific SME component. As a result the oscillation probability is split into two annular regions separated by two null directions, seen clearly in the left panels of Figure\,\ref{fig: variation day signal Toulon}. \\
When the core-crossing, longest atmospheric neutrino baselines align with the field, the oscillations are enhanced, resulting in a second oscillation maximum being visible in addition to the first for the given parameter choice, as seen in the right panels of Figure\,\ref{fig: variation day signal Toulon}.

\begin{figure*}[p]
    \includegraphics[trim={8.5cm 2cm 21cm 2cm},clip,width=\linewidth]{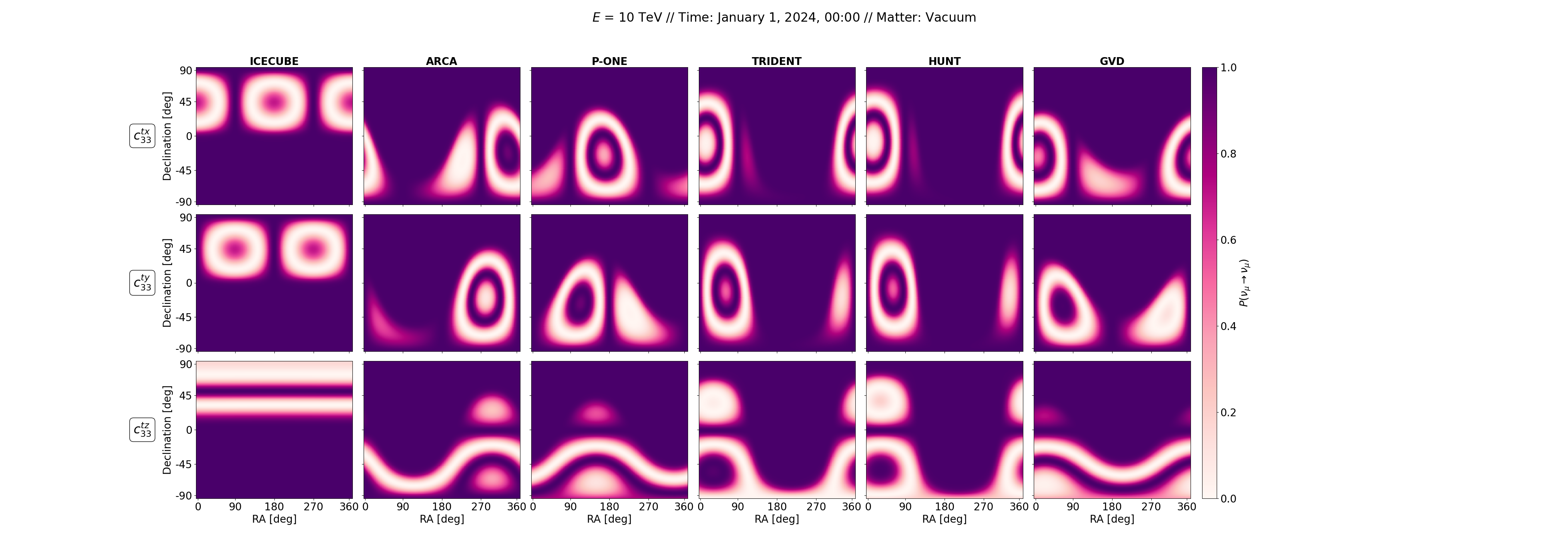}
    \caption{\textbf{LIV-induced vacuum $\nu_\mu$ oscillation probabilities for six large-scale neutrino detectors.} We show $P(\nu_\mu\rightarrow\nu_\mu)$ for atmospheric $\nu_\mu$ with  $E=10$\,TeV as a function of right ascension and declination. Each row shows the oscillation pattern for a single spatial SME component
    $(c^{(4)tx}_L)_{33},(c^{(4)ty}_L)_{33}, (c^{(4)tz}_L)_{33}$, with magnitude $7.9\cdot 10^{-27}$\, (being non-zero at a time).
    Each column corresponds to a different existing or proposed neutrino telescope: IceCube, KM3NeT/ARCA, P-ONE, TRIDENT, HUNT, and GVD.
    Annular oscillation regions appear, separated by null directions where the propagation is orthogonal to the field. The morphology varies between detectors due to latitude dependent mapping between RA/DEC and zenith angle (and therefore baseline). Additionally, because the plot corresponds to a fixed time (Jan.\ 1, 2024, 00:00 UTC), detectors at different longitudes appear at different phases of their sidereal modulation.}
    \label{fig: Sidereal dependence all detectors}
\end{figure*}

While we showed in Figure\,\ref{fig: variation day signal Toulon} the oscillation probabilities for a specific orientation, in practice the background field could have any orientation. In  Figure\,\ref{fig: Sidereal dependence all detectors} we therefore display the signatures for the three spatial dipolar components of $(\widehat{c}_L^{(4)t\beta})_{33}$ across six existing and proposed large scale neutrino detectors designed to measure high-energy neutrinos: IceCube\,\cite{IceCube:2016zyt}, KM3NeT/ARCA\,\cite{KM3Net:2016zxf}, P-ONE\,\cite{agostiniPacificOceanNeutrino2020}, TRIDENT\,\cite{yeMulticubickilometreNeutrinoTelescope2024}, HUNT\,\cite{huangProposalHighEnergy2023}, and GVD\,\cite{belolaptikovBaikalUnderwaterNeutrino1997}. Although the latitude of each detector alters the geometric mapping of the baseline, the fundamental anisotropic LIV signature, which consists of annular oscillation regions separated by null directions, remains universal. 

Finally, we note that the dimension-4 operator $\widehat{c}_L^{(4)}$ also includes additional multipoles. The spherical harmonic decomposition of $\widehat{c}_L^{(4)}$ in Eq.\,\eqref{eq: Spherical decomposition aL3} includes a monopole ($l=0$) term and quadrupole ($l=2$) moments. \\
The monopole moment is isotropic. It induces neutrino oscillations that depend only on energy and baseline, without any intrinsic direction dependence in the Sun-centered frame. \\
In contrast, quadrupole terms introduce additional angular structure, creating more complex patterns with higher frequency modulations in RA and DEC and additional null directions. The phenomenology of the quadrupole moments is visually distinct and is compared to the dipole moments in Appendix\,\ref{appendix1} for  KM3NeT/ARCA as a reference. 
The decomposition of higher dimensional coefficients of the SME includes even higher order multipoles. Given this rich phenomenology future analysis could benefit from generalized searches for directional modulations to identify anisotropic anomalies without relying on specific operator templates. 

The directional analysis presented so far does not include a realistic matter potential. In reality, neutrinos propagate through a varying matter potential, which can interfere with the LIV-induced neutrino oscillations. In the next section, we examine how this interplay can alter the oscillation probabilities.

\begin{figure*}[p]
    \includegraphics[trim={8.5cm 2cm 21cm 2cm},clip,width=\linewidth]{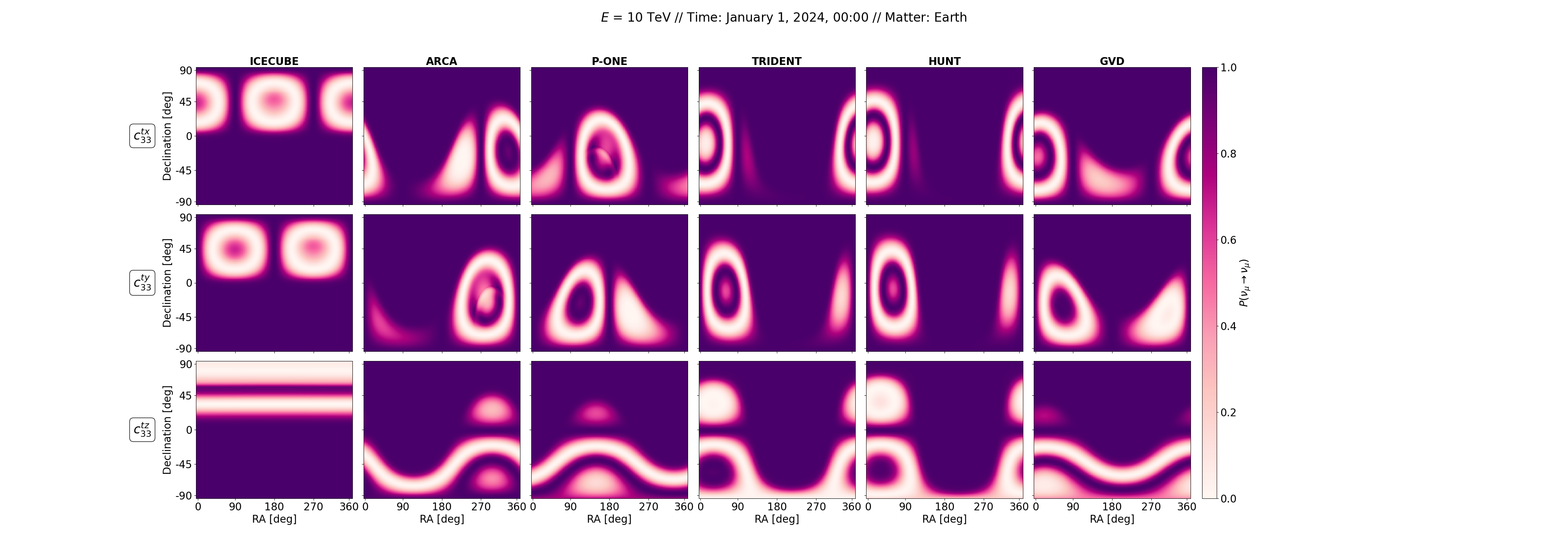}
    \caption{\textbf{LIV-induced $\nu_\mu$ oscillation probabilities for six large-scale neutrino detectors in matter.} We show $P(\nu_\mu\rightarrow\nu_\mu)$ for atmospheric $\nu_\mu$ with  $E=10$\,TeV as a function of right ascension and declination assuming a realistic Earth model. Each row shows the oscillation pattern for a single spatial SME component 
    $(c^{(4)tx}_L)_{33},(c^{(4)ty}_L)_{33}, (c^{(4)tz}_L)_{33}$, with magnitude $7.9\cdot 10^{-27}$\, (being non-zero at a time).
    Columns correspond to IceCube, KM3NeT/ARCA, P-ONE, TRIDENT, HUNT, and GVD.In contrast to the vacuum case, the patterns here are modified in dense regions in Earth's interior and by density transitions, particularly near mantle–core boundaries.}
    \label{fig: Sidereal dependence all detectors matter}
\end{figure*}

\section{Interplay of LIV and Matter}\label{sec4}

Understanding the modification of neutrino oscillations in the presence of matter due to the Mikheyev-Smirnov-Wolfenstein (MSW) effect\,\cite{Wolfenstein:1977ue,Mikheev:1986if,SNO:2001kpb} was crucial to understanding the reduced solar $\nu_e$ flux observed, and solar neutrino oscillations\,\cite{SNO:2001kpb}. Similarly, the matter potential of the Earth can enhance neutrino oscillations through parametric resonance\,\cite{Petcov:1998su,Akhmedov:1998ui}. 

In the presence of LIV, this picture becomes richer. While the matter potential $V$ in the Hamiltonian in Eq.\,\eqref{eq: full Hamiltonian} is energy-independent, the contribution of SME coefficients of dimension $d>3$ to the Hamiltonian grows with energy $\propto E^{d-3}$. Consequently, there exists a specific regime where the LIV contribution becomes comparable in magnitude to the matter potential, $H_{\text{SME}}\sim V$. In this crossover regime, the two effects compete directly and they can constructively interfere to generate novel resonances or destructively interfere to suppress LIV-induced oscillations.

This interplay produces distinct spectral and directional distortions absent in vacuum, as illustrated qualitatively in Figure\,\ref{fig: Sidereal dependence all detectors matter}. 
The oscillation patterns in matter differ noticeably from the vacuum case in Figure\,\ref{fig: Sidereal dependence all detectors}, 
especially inside dense regions and at Earth layer boundaries. 
In the remainder of this section, we develop intuition for why these effects occur 
and how they manifest observationally.

The following section is split into three parts. 
First, we analyze how matter impacts LIV-induced oscillations in simplified setups and map this to realistic Earth models.
Second, we examine how matter breaks the symmetry between neutrinos and antineutrinos leading to separate signatures. 
Finally, we discuss how enhanced $\nu_\tau$ production inside the Earth can amplify $\tau$ regeneration, creating a distinct flux signature at neutrino telescopes.

\subsection{Resonant Enhancement of LIV-induced Neutrino Oscillations in Matter}\label{subsub1}
%
%
%
Matter affects the propagation of neutrinos due to the electrons contained in it. The matter potential in Eq.\,\eqref{eq: full Hamiltonian} written explicitly in the flavor basis is given by 
\begin{align}
    V=\pm \sqrt{2}G_F\mqty(N_e & 0&0\\0 & 0&0\\0 & 0&0),
    \label{eq: matter potential}
\end{align}
where $G_F$ is the Fermi constant and $N_e$ is the electron density\,\cite{kuoNeutrinoOscillationsMatter1989}. The interplay of this potential with the SME Hamiltonian in Eq.\,\eqref{eq: HLIV} gives rise to the clear differences in the oscillation patterns in Figure\,\ref{fig: Sidereal dependence all detectors matter} compared to the vacuum case in Figure\,\ref{fig: Sidereal dependence all detectors}. These differences are evident in regions with a high electron density $N_e$ (compare the IceCube $(c^{(4)ty}_L)_{33}$ panels with each other) or at sharp density transitions (compare the ARCA $(c^{(4)tx}_L)_{33}$ panels with each other). 
\begin{figure}[t]
    \includegraphics[width=0.9\linewidth]{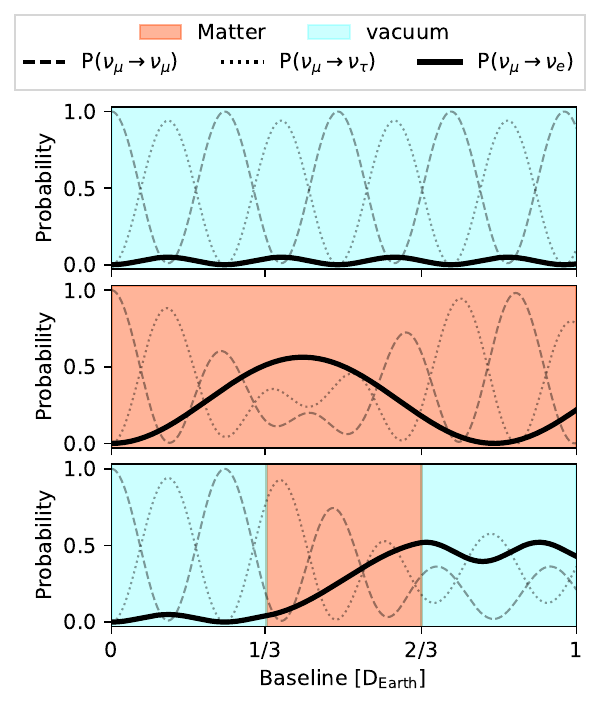}
  \caption{\textbf{Enhancement of LIV-induced oscillations in matter: comparison of vacuum, constant density, and layered density models.} Evolution of the oscillation probabilities, $P(\nu_\mu\rightarrow\nu_x)$, $x\in \{e,\mu,\tau\}$, for baselines of the order of Earth's diameter $D_\text{Earth}$ in a) vacuum, b) constant electron density matter, c) a layered density profile. The constant density case shows a strong enhancement of $P(\nu_\mu\leftrightarrow\nu_e)$ relative to vacuum, while the layered profile illustrates how flavor conversion can be fixed through density transitions. The matter density is $\rho=10$\,g/cm$^3$ with the electron fraction $f=0.5$ and the SME component is set to $(a^{(3)}_L)_{33}=4\cdot 10^{-22}~$\,GeV.}
  \label{fig: vacuum vs matter osc.prob}
\end{figure}
\begin{figure*}[t]
    \centering
    \includegraphics[width=0.8\linewidth,trim={0cm 0cm 0cm 0cm},clip]{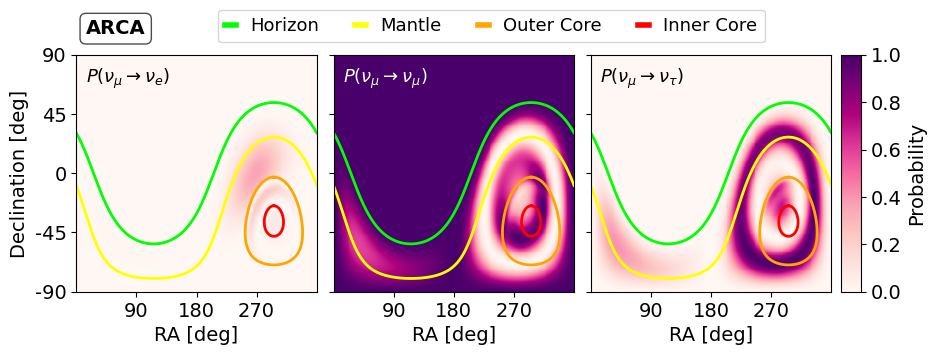}
    \caption{\textbf{Resonance enhancement of the $\nu_\tau$ oscillation probability inside the Earth's core.} We plot the oscillation probabilities for $10$\,TeV atmospheric $\nu_\mu$ at the KM3NeT/ARCA detector as a function of right ascension and declination. For an anisotropic LIV field $c^{ty}_{33}=7.9\cdot 10^{-27}~$, we observe an increase of the $P(\nu_\mu\rightarrow\nu_\tau)$ oscillation probability for neutrinos traveling through the Earth's core, while for neutrinos crossing the mantle $P(\nu_\mu\rightarrow\nu_e)$ is enhanced for certain directions.}
    \label{fig: anisotropic neutrino oscillation probabilities}
\end{figure*}

To isolate the physical mechanism driving these patterns, we first examine a simplified 1D system in Figure\,\ref{fig: vacuum vs matter osc.prob}. The top panel shows the vacuum evolution of a pure $\nu_\mu$ propagating over a distance equal in length to Earth's diameter antiparallel to a LIV field that only affects the third mass eigenstate. This is the same setup as in the previous section. Similar to standard mass-induced atmospheric neutrino oscillations, we observe in the vacuum case almost maximal $\nu_\mu\leftrightarrow\nu_\tau$ oscillations with negligible $\nu_\mu\leftrightarrow\nu_e$ oscillations. In the middle panel, we introduce a medium of constant density. The matter potential generates an effective mixing between the flavor states, enabling a resonant enhancement of the $P(\nu_\mu\leftrightarrow\nu_e)$. This resonant enhancement of neutrino oscillations occurs when the vacuum oscillation length becomes comparable to the refraction length $l_0=\sqrt{2}\pi/(G_FN_e)$\cite{smirnovMSWEffectMatter2005}. 

In reality, however, the Earth is not a constant medium, but rather comprises multiple layers of varying density.
The bottom panel of Figure\,\ref{fig: vacuum vs matter osc.prob} more closely reflects reality with a sandwich model. One third of the baseline in the middle contains matter of constant density (same density as in the middle panel) sandwiched in between two vacuum layers. We initially observe a small $P(\nu_\mu\leftrightarrow\nu_e)$ as in the vacuum case. But upon entering the constant density block at $1/3$ of the baseline, the change in the effective Hamiltonian activates the $\nu_\mu \to \nu_e$ conversion. Crucially, when the neutrino exits the matter and returns to vacuum, the probability $P(\nu_\mu\leftrightarrow\nu_e)$ does not revert to zero. Instead, the flavor composition acquired inside the matter is "frozen in" because the vacuum mixing angle for this channel is negligible. \\
This demonstrates that discontinuous density profiles can act as "flavor converters," fixing the neutrino state into a configuration that would be forbidden by vacuum physics alone.

In realistic Earth density profiles, these interference effects persist and additionally include strongly directional signatures. In Figure\,\ref{fig: anisotropic neutrino oscillation probabilities} we show the $P(\nu_\mu\rightarrow\nu_e)$, $P(\nu_\mu\rightarrow\nu_\mu)$, and $P(\nu_\mu\rightarrow\nu_\tau)$ oscillation probabilities of $10$\,TeV atmospheric $\nu_\mu$ at KM3NeT/ARCA for a constant LIV background field given by $(c^{(4)ty}_L)_{33}=7.9\cdot 10^{-27}~$. The Earth is modeled through the Preliminary Reference Earth Model (PREM)\,\cite{Dziewonski:1981xy}, which approximates the Earth by 200 layers of constant density. We note that a simplified Earth model with 5 layers reproduces the oscillation features with high fidelity. This indicates that the resonance phenomena are driven by the macroscopic density contrasts between the Earth's main layers (inner/outer core, mantle), rather than by fine-grained density gradients. This confirms that the sharp boundaries observed in the figure are physical consequences of the transitions between the outer and inner core, and the outer core and the mantle, whose boundaries we plot as colored lines over the  oscillation probabilities for reference. 

The overall behavior is consistent with the discussion for Figure \ref{fig: vacuum vs matter osc.prob}. We observe the resonant enhancement of the $P(\nu_\mu\rightarrow\nu_e)$ probability for baselines through the denser layer of the Earth's mantle. However, this effect does not occur symmetrically for all neutrino paths of the same length crossing the mantle. This enhancement is contained to regions around a declination of DEC$\approx0^\circ$ due to anisotropy of the LIV field.\\
We can also see a clear enhancement of the $P(\nu_\mu\rightarrow\nu_\tau)$ inside the Earth's outer core that is contained by the edge of the outer core clearly showing the impact of the different density layers on LIV-induced oscillations. This shows that the LIV-matter interplay can affect the flavor composition of atmospheric neutrinos in the energy range of current high-energy neutrino telescopes for currently unconstrained SME parameter values.

The findings presented here for a specific renormalizable operator also apply to higher-dimensional SME terms, because these operators scale even more strongly with energy and a crossover regime where $H_{\text{SME}}\sim V$ is inevitable. While we focused in this subsection on the interference between LIV and the Earth's interior, this interplay is a universal feature of propagation in any non-vacuum environment. We expect similar interference signatures to emerge for astrophysical neutrinos traversing varying electron densities, potentially altering the flavor composition of the associated neutrino flux.

\subsection{Directional Neutrino–Antineutrino Splitting}\label{subsub2}
%
We have so far focused on neutrino oscillation probabilities, reflecting the difficulty current large-scale neutrino telescopes have at separating neutrinos from antineutrinos on an event-by-event basis\,\cite{collaborationMeasurementsUsingInelasticity2019}. However, the interplay of LIV and matter effects breaks the symmetry between neutrinos and antineutrinos in the form of direction dependent signatures that could be used for statistical separation.

The SME Hamiltonian in Eq.\,\eqref{eq: HLIV} includes both CPT-even and CPT-odd operators\,\cite{Kostelecky:2003LIVCPTVNeutrinos}. CPT-odd operators flip sign between neutrinos and antineutrinos, whereas CPT-even terms do not. However, if the SME term dominates the evolution, neither case yields an observable $\nu$-$\Bar{\nu}$ difference because the oscillation probabilities are invariant under an overall sign flip\footnote{A directional $\nu$–$\bar{\nu}$ asymmetry may arise even in vacuum when both CPT-even and CPT-odd operators are present simultaneously, provided their contributions are comparable and sufficiently large to induce oscillations. This work focuses instead on the interplay between LIV and matter effects, where the asymmetry originates from the opposite sign of the matter potential for antineutrinos\,\cite{Wolfenstein:1977ue}.}. However, matter breaks $\nu$-$\Bar{\nu}$ symmetry, when the contribution of the SME term and the matter term to the Hamiltonian in Eq.\,\eqref{eq: full Hamiltonian} are comparable.

\begin{figure}
    \centering
    \includegraphics[width=1.1\linewidth]{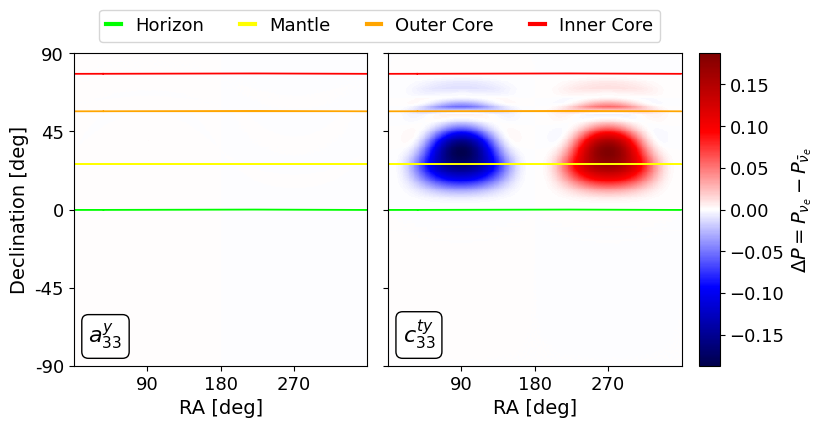}
    \caption{ \textbf{Directional neutrino–antineutrino separation from the interplay of matter effects and CPT-even and CPT-odd LIV operators.} 
    This figure shows the residuals between the oscillation probabilities $P(\nu_\mu\rightarrow\nu_e)$ (red) and $P(\Bar{\nu}_\mu\rightarrow\Bar{\nu}_e)$ (blue) at IceCube as a function of RA/DEC for a realistic Earth model. The left panel shows the CPT-odd operator $(a^{(3)y}_L)_{33}= 1.58\cdot 10^{-13}$ eV, showing no separation, while the right panel shows the CPT-even operator $(c^{(4)ty}_L)_{33}= 7.9\cdot 10^{-27}$ at an energy of $E=10$ TeV, showing a significant neutrino-antineutrino separation.
    The figure highlights how the sign flip of the matter potential, and CPT-odd operators for antineutrinos, cancel out, whilst for CPT-even operators, it results in a clear separation between neutrinos and antineutrinos, characterized by an $180^\circ$ reversal of the directional anisotropy.
    }
    \label{fig: Neutrino-Antineutrino Separation}
\end{figure}

The right panel of Figure\,\ref{fig: Neutrino-Antineutrino Separation} illustrates this for a CPT-even operator. For neutrinos the matter enhanced $P(\nu_\mu\rightarrow\nu_e)$ oscillation probability occurs for neutrinos propagating antiparallel to the LIV field, while for antineutrinos the matter enhanced $P(\Bar{\nu}_\mu\rightarrow\Bar{\nu}_e)$ oscillation probability occurs for the opposite direction when the antineutrinos propagate parallel to the LIV field. Although the oscillation probabilities are shown for the IceCube detector and $(c^{(4)ty}_L)_{33}= 7.9\cdot 10^{-27}$, the effect is generic for anisotropic LIV fields in the presence of a sizable matter contribution. The directional $\nu$–$\bar{\nu}$ separation occurs because for antineutrinos the matter potential in Eq.\eqref{eq: matter potential} changes sign, while the CPT-even operator does not. This means that the condition that leads to a resonant enhancement of the $P(\nu_\mu\rightarrow\nu_e)$ oscillation probability for neutrinos propagating antiparallel to the LIV field, occurs for antineutrinos in the parallel direction due to the direction-dependent contribution of the anisotropic LIV field. \\

For CPT-odd SME operators, no observable directional difference between neutrino and antineutrino oscillation probabilities is expected because both the matter potential and the CPT-odd operators change sign for antineutrinos and the overall sign change does not affect neutrino oscillation probabilities. This is supported by the left panel of Figure\,\ref{fig: Neutrino-Antineutrino Separation}, where the neutrino and antineutrino oscillation probability $P(\nu_\mu\rightarrow\nu_e)$ and $P(\Bar{\nu}_\mu\rightarrow\Bar{\nu}_e)$ show no separation in the probability residuals for the CPT-odd operator $(a^{(3)y}_L)_{33}$. 

The same reasoning extends to higher dimensional SME operators. This emphasizes that the interplay of CPT-even LIV operators with matter can provide a physical mechanism to distinguish neutrinos and antineutrinos through their arrival directions.

\subsection{LIV-Enhanced $\tau$ Regeneration in the Earth}\label{subsub3}
In Subsection \ref{subsub1} we observed that LIV can induce sizable $\nu_\mu\rightarrow\nu_\tau$ oscillations inside the Earth, particularly for neutrino baselines crossing the Earth's core, see Figure\,\ref{fig: anisotropic neutrino oscillation probabilities}. Here we discuss how this can lead to significant enhancements of $\nu_\tau$ regeneration effects.

\begin{figure}
    \centering
    \includegraphics[width=0.75\linewidth,trim={0.0cm
    0.0cm 0.0cm 0.0cm},clip]{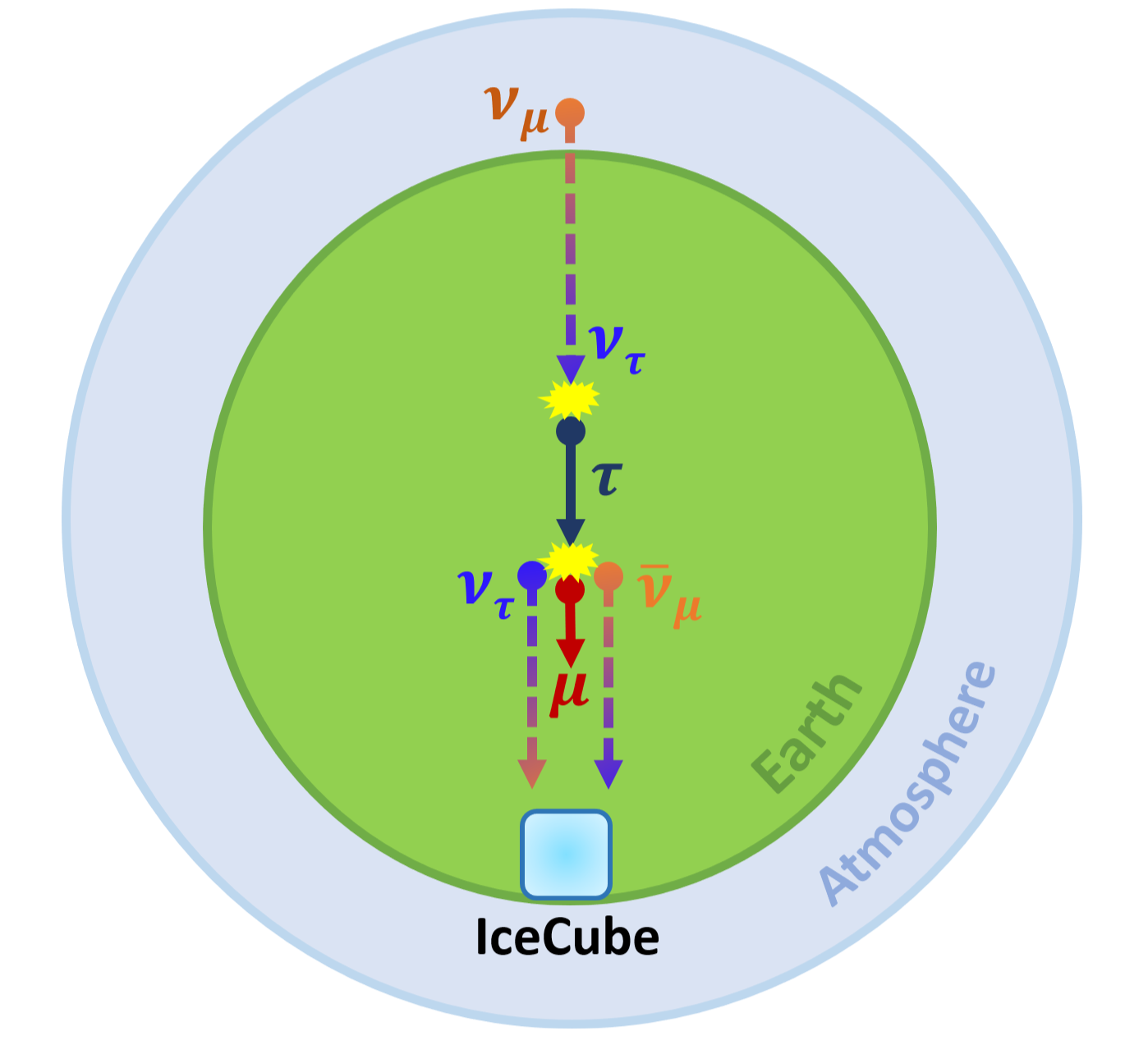}
    \caption{\textbf{Schematic Illustration of LIV-enhanced $\tau$ regeneration.} LIV-induced $\nu_\mu\rightarrow\nu_\tau$ oscillations inside the Earth can convert $\nu_\mu$ dominated neutrino fluxes into a $\nu_\tau$ flux, which can undergo $\tau$ regeneration. This effectively can convert a high-energy neutrino flux into a lower energy neutrino flux.}
    \label{fig: tau_regen_diagram}
\end{figure}

At TeV energies and above, the Earth is no longer transparent to neutrinos. Neutrinos crossing the Earth can have a significant probability of interacting before reaching a detector, depleting the detected neutrino flux\cite{cooper-sarkarHighEnergyNeutrino2011,IceCube:2017roe}. However, $\nu_\tau$ are unique. When a $\nu_\tau$ undergoes a charged-current interaction in matter it produces a short lived $\tau$ lepton that decays rapidly, producing a secondary and lower energy $\nu_\tau$ (and potentially $\nu_e$/ $\nu_\mu$ secondaries) \cite{halzenTauNeutrinoAppearance1998}. This process is known as $\tau$ regeneration. The cycle can repeat multiple times and can partially restore the flux of Earth-crossing $\nu_\tau$, but shifts it towards lower energies. 

For atmospheric neutrinos in the standard mass-induced oscillations regime, the regeneration effect is limited because the atmospheric flux mainly consists of $\nu_\mu$ and $\nu_e$, and only a small fraction converts to $\nu_\tau$ before being absorbed. For LIV-induced oscillations this picture changes drastically. LIV-induced neutrino oscillations can enhance the $\nu_\mu \to \nu_\tau$ transition at high energies, providing a mechanism that injects a high-energy $\nu_\tau$-rich flux inside the Earth. We note that this enhanced $\nu_\tau$ injection mechanism is not exclusive to anisotropic operators: any LIV operator—isotropic or anisotropic—that produces sizable $\nu_\mu\rightarrow\nu_\tau$ conversion at TeV energies will trigger the same amplification of the $\tau$ regeneration cycle. The anisotropic nature of the operators studied here introduces a direction dependence to the amplitude of the effect: the enhancement is largest when neutrinos propagate parallel to the LIV background field, and vanishes for orthogonal directions. These $\nu_\tau$ can undergo repeated $\tau$ regeneration cycles \footnote{We illustrate this phenomenon schematically in Figure\,\ref{fig: tau_regen_diagram}, where LIV results in the conversion of high-energy $\nu_\mu$ into $\nu_\tau$, which undergo $\tau$ regeneration (we specifically show the leptonic tau decay $\tau\rightarrow\mu\Bar{\nu}_\mu\nu_\tau$ that also produces a $\nu_\mu$ secondary) and survive to reach the detector. In the absence of LIV, the $\nu_\mu$ would most-likely be absorbed and never reach the detector volume.}.

\begin{figure}
    \includegraphics[width=0.95\linewidth,trim={0.02cm 0.02cm 0.05cm 0cm},clip]{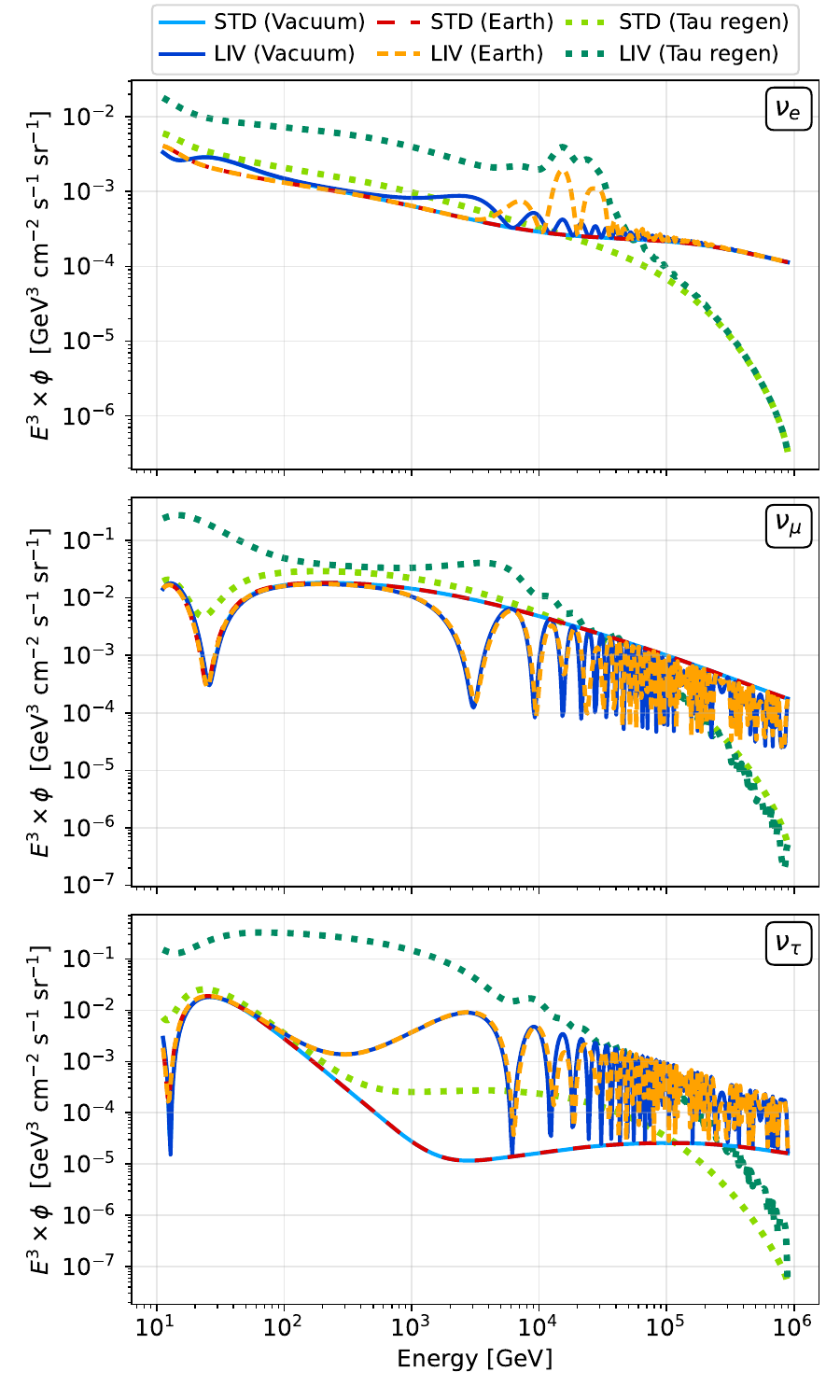}
    \caption{\textbf{Flavor-separated atmospheric neutrino fluxes for core-crossing trajectories.} We compare the final $\nu_e$, $\nu_\mu$, and $\nu_\tau$ fluxes for standard oscillations and LIV-induced oscillations ($c^{(4)tz}_{33} = 7.9\cdot 10^{-27}$). For each case, we show the progression from vacuum propagation (solid) to the inclusion of matter effects (dashed) and finally matter effects including $\tau$ regeneration (dotted). The amplification of the $\nu_\tau$ flux (dark green dotted line) illustrates the flux enhancement driven by the injection of high-energy $\nu_\tau$ through LIV-induced oscillations into the $\tau$ regeneration cycle.}
    \label{fig: flavor separated flux}
\end{figure}
\begin{figure}
    \includegraphics[width=0.95\linewidth,trim={0.4cm 0.2cm 2.9cm 1cm},clip]{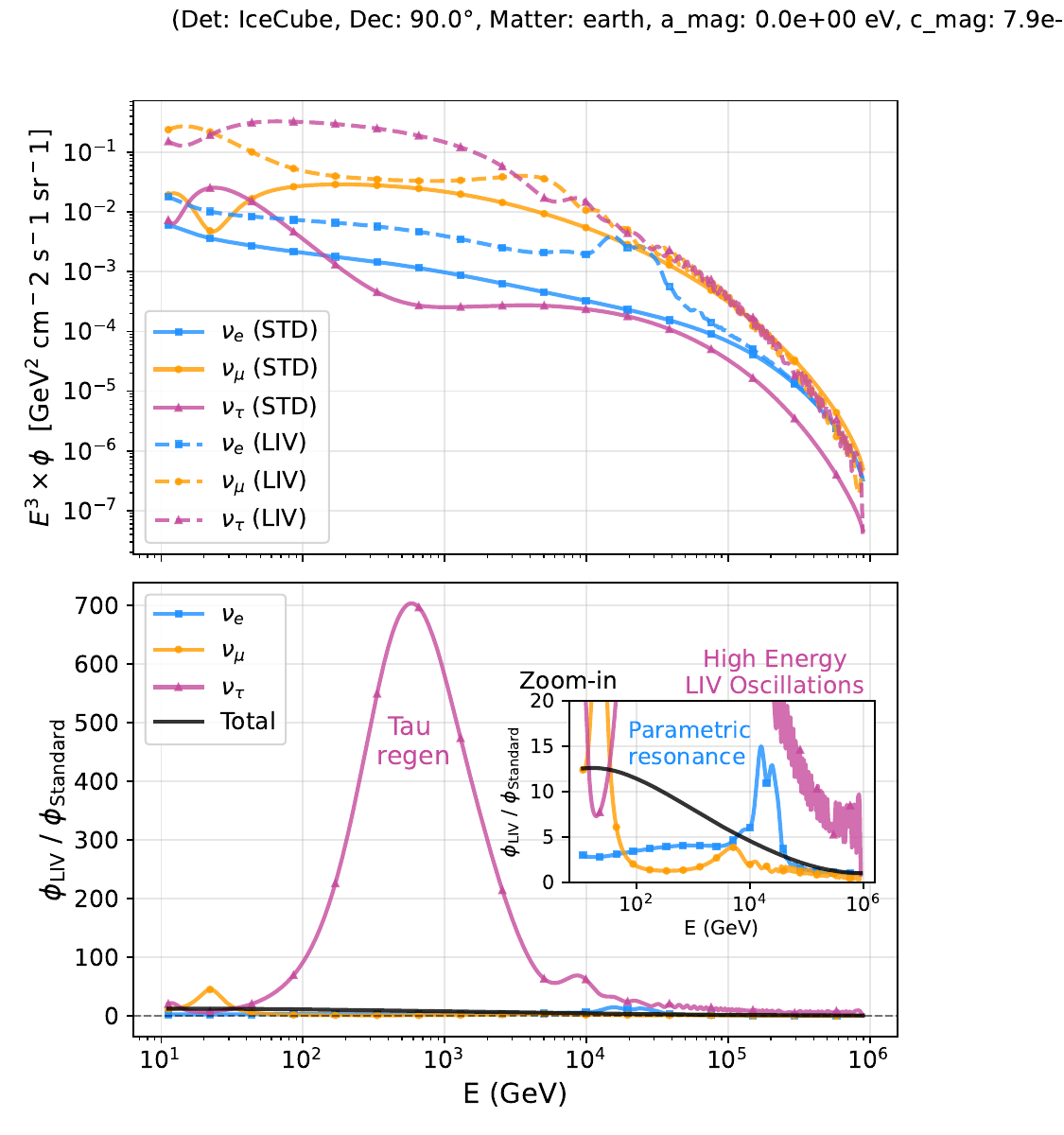}
    \caption{\textbf{Atmospheric flux ratios illustrating the $\tau$ regeneration signature.} We show the ratio of the final fluxes with and without LIV ($c^{(4)tz}_{33} = 7.9\cdot 10^{-27}$) for core-crossing neutrinos ($\cos{\theta_z}=-1$). Curves are shown for $\nu_e$ (blue, squares), $\nu_\mu$ (yellow, dots), and $\nu_\tau$ (pink, triangles), along with the total flux ratio (black, solid). The LIV field is assumed to be antiparallel to the neutrino propagation. The synergy between high-energy $\nu_\mu \to \nu_\tau$ conversion and subsequent regeneration creates a factor of 100 enhancement in the $\nu_\tau$ flux at $10^3$\,GeV.}
    \label{fig: tau regeneration atmospheric neutrino flux}
\end{figure}

We illustrate the impact of this on the propagated Earth-crossing neutrino fluxes in Figure\,\ref{fig: flavor separated flux}, exemplarily for the IceCube detector location. For each flavor, we compare six scenarios: standard and LIV cases across vacuum, matter without $\tau$ regeneration, and matter with $\tau$ regeneration. The key observation is the strong divergence in the $\nu_\tau$ panel, where for the LIV case with matter effects and $\tau$ regeneration (dark green dotted line) the flux is several orders of magnitude larger than in the standard case (light green dotted line). Also, comparing the dark green dotted line to the LIV case in vacuum (dark blue solid line), shows that the interplay of LIV with matter is significant. LIV searches that neglect matter effects and $\tau$ regeneration would fail to predict these signatures.\\
Additionally, the full LIV case shows a notable increase in $\nu_e$ and $\nu_\mu$ fluxes at lower energies ($\sim 10^2$\,GeV) compared to the vacuum cases. High-energy neutrinos in the presence of LIV are not just lost to absorption they regenerate at lower energies as a mixture of all three flavors via leptonic $\tau$ decay channels $\tau\rightarrow l\Bar{\nu}_l\nu_\tau$ and LIV-induced oscillations.

This spectral redistribution is quantified in Figure\,\ref{fig: tau regeneration atmospheric neutrino flux}. 
In the top panel we compare the expected atmospheric neutrino fluxes with and without LIV for core-crossing trajectories including matter effects and $\tau$ regeneration. The $\nu_\tau$ flux component (pink dashed line with triangles) is dominant between $10^2$ and $10^4$\,GeV, where usually the $\nu_\mu$ component would dominate. \\
In the bottom panel, the flux ratios reveal that LIV-enhanced $\tau$ regeneration leads to a higher total neutrino flux (black solid line) at lower energies ($10^2 -10^4$\,GeV). 

Two specific features in the flux ratios merit special attention. First, the LIV-enhanced $\tau$ regeneration leads to $\mathcal{O}(100)\times$ increased $\nu_\tau$ flux near $10^3$\,GeV. Second, there is a visible bump in the $\nu_e$ ratio around $10^4$\,GeV. This is a direct manifestation of the LIV matter interplay discussed in Subsection \ref{subsub1}. A resonant conversion produces an amplified $\nu_e$ flux for this specific field configuration.

To produce the results shown in Figures\,\ref{fig: flavor separated flux} and\,\ref{fig: tau regeneration atmospheric neutrino flux}, atmospheric neutrino fluxes were generated using the {\ttfamily MCEq} package~\cite{Fedynitch:2015zma}, employing the \textit{HillasGaisser2012}\,\cite{gaisserSpectrumCosmicrayNucleons2012} primary flux model and the \textit{SIBYLL23C}\,\cite{riehnHadronicInteractionModel2018} hadronic interaction model. These fluxes were then propagated through a 200-layer \textit{PREM} Earth model~\cite{Dziewonski:1981xy} using the {\ttfamily nuSQuIDS} software~\cite{Arguelles:2021twb}. Our analysis accounts for the impact of LIV on the neutrino propagation and flavor evolution, but not potential LIV corrections to particle interaction kinematics. Such effects could, in principle, introduce a high-energy decay threshold to the $\tau$ lepton\footnote{For a comprehensive discussion of threshold effects, see Sec. VII B in\,\cite{kosteleckyNeutrinosLorentzviolatingOperators2012}. For an application to the explanation of a lack of Glashow resonance events at IceCube, see\,\cite{Tomar:2015GlashovResonance}.}, but given the large $\tau$ rest mass relative to the lighter decay products ($\pi$, $\mu$, and $e$), we do not expect threshold effects to invalidate the $\tau$ regeneration signatures for the energy ranges and LIV operator strengths considered in this work.

The significance of LIV-enhanced $\tau$ regeneration and resonant flavor conversions in matter including factor 10 to factor 100 enhancements, suggest the LIV matter interplay could have strong impact on LIV searches with neutrinos. While this analysis focuses on core-crossing neutrinos $\cos\theta_z=-1$ to maximize the observable effect, the qualitative features are universal for any baseline traversing dense matter. Neglecting the interplay of LIV with matter potentials of dense media traversed by neutrinos could lead to missing the most distinct signatures of spacetime symmetry violation.

Regarding the experimental observability of these effects, we note that the LIV-enhanced $\tau$ regeneration peaks in the energy range $10^2$–$10^4$\,GeV, which overlaps directly with the sensitivity range of current large-volume neutrino telescopes such as IceCube~\cite{IceCube:2016zyt} and KM3NeT~\cite{KM3Net:2016zxf}. The directional $\nu$–$\bar{\nu}$ splitting, which manifests as an $180^\circ$ reversal of the enhancement pattern, could in principle be probed by statistically separating neutrino and antineutrino events via inelasticity measurements~\cite{collaborationMeasurementsUsingInelasticity2019}. A full sensitivity study, including detector response, backgrounds, and statistical inference, is beyond the scope of this work. However, the factor $\sim 100$ flux enhancement in the $\nu_\tau$ channel at $\sim 10^3$\,GeV suggests that even modest $\nu_\tau$ identification efficiency, of the kind achievable at IceCube or at KM3NeT/ARCA, would yield sensitivity to these signatures. We therefore expect these effects to be phenomenologically relevant for next-generation LIV searches at large-scale neutrino telescopes.

\section{Summary}\label{ch5}
This work demonstrates that the interplay of Lorentz-invariance violating operators and the Earth's matter potential produces distinct signatures observable in current and future large-scale neutrino observatories. For LIV operators with positive energy scaling, we identify an inevitable high-energy regime where the LIV contribution to the Hamiltonian becomes comparable to the matter potential. This competition results in a rich phenomenology of direction-dependent resonant enhancements and physical neutrino-antineutrino splitting. Furthermore, we propose LIV-enhanced $\tau$ regeneration as a novel flux signature. Driven by the conversion of atmospheric neutrinos to $\nu_\tau$ within the Earth, this process could generate a significant excess in the expected number of $\nu_\tau$ events, and a spectral bump in the total Earth-crossing atmospheric neutrino flux.  These results highlight the importance of matter effects for LIV searches and present unique signatures, opening a new window for discovery.

\vspace{0.2cm}

\begin{acknowledgments}
This project has received funding from the European Union’s Horizon Europe research and innovation programme under the Marie Sklodowska-Curie grant agreement No 101168829, Challenging AI with Challenges from Physics: How to solve fundamental problems in Physics by AI and
vice versa (AIPHY).  T.S. and D.J.K. acknowledge support from the Carlsberg Foundation (No.~117238). J.I.-N. performed part of this work as a visiting researcher supported by the Erasmus+ program of the European Union.
\end{acknowledgments}
\newpage

\bibliographystyle{SciPost-bibstyle} 

\bibliography{janni}

\newpage
\appendix
\section{Dipole and Quadrupole Components of LIV operators}\label{appendix1}
We specifically show the dipole and quadrupole moment of $(c_L^{(4)\alpha\beta})_{33}$ here. The isotropic monopole is omitted as it lacks directional structure.
\begin{figure}[h!]
    \centering
    \includegraphics[trim={2.8cm 1.8cm 6cm 3cm},clip,width=\linewidth]{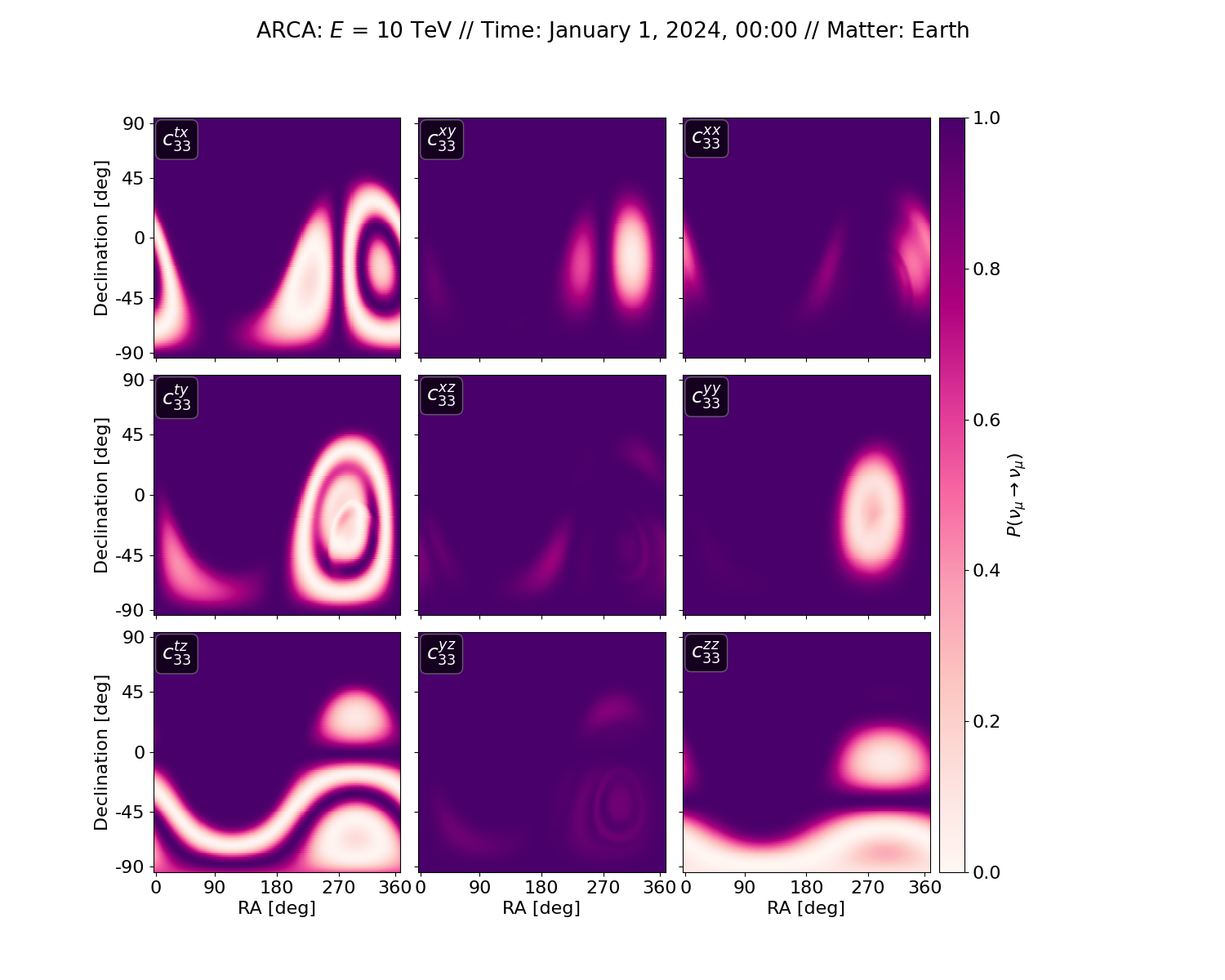}
    \caption{\textbf{Comparison of Dipole and Quadrupole LIV signatures at KM3NeT/ARCA.}  We show $P(\nu_\mu\rightarrow\nu_\mu)$ for atmospheric $\nu_\mu$ with  $E=10$\,TeV as a function of right ascension and declination assuming a realistic Earth model. The left column displays the dipolar ($l=1$) signatures arising from $(c_{L}^{(4)t\beta})_{33}$. The center and right columns display the quadrupole ($l=2$) signatures arising from the spatial components, which exhibit higher-frequency angular modulations and a more complex topology of null bands. All plots are shown for an amplitude $7.9\cdot 10^{-27}$.}
    \label{fig:SME monopole and quadrupole of c}
\end{figure}
\end{document}